\documentclass[
reprint,
showkeys,
longbibliography,
nofootinbib,
superscriptaddress,
aps,
notitlepage
]{revtex4-1}
\usepackage[utf8]{inputenc}

\usepackage{amsmath}
\usepackage{amssymb}
\usepackage{bm}
\usepackage{enumitem}
\usepackage[acronym]{glossaries}
\usepackage{graphicx}
\usepackage[colorlinks,unicode]{hyperref}
\usepackage[capitalise]{cleveref}  % must come after hyperref
\usepackage{mathtools}
\usepackage{multirow}
\usepackage{physics}
\usepackage{siunitx}
\usepackage{ulem}
\usepackage[dvipsnames]{xcolor}

\makeglossaries
\newacronym{dm}{DM}{dark matter}
\newacronym{ce}{CE}{Cosmic Explorer}
\newacronym{et}{ET}{Einstein Telescope}
\newacronym{snr}{SNR}{signal-to-noise ratio}

\makeatletter
\renewcommand\onecolumngrid{%
\do@columngrid{one}{\@ne}%
\def\set@footnotewidth{\onecolumngrid}%
\def\footnoterule{\kern-6pt\hrule width 1.5in\kern6pt}%
}
\makeatother

\newcommand\myshade{80}
\colorlet{mylinkcolor}{ForestGreen}
\colorlet{mycitecolor}{Red}
\colorlet{myurlcolor}{violet}

\hypersetup{
  linkcolor  = mylinkcolor!\myshade!black,
  citecolor  = mycitecolor!\myshade!black,
  urlcolor   = myurlcolor!\myshade!black,
  colorlinks = true
}

\DeclareSIUnit\solarmass{\ensuremath{\mathrm{M}_\odot}}
\DeclareSIUnit\parsec{pc}
\DeclareSIUnit\year{yr}

\newcommand{\GRAPPA}{Gravitation Astroparticle Physics Amsterdam (GRAPPA),\\ Institute for Theoretical Physics Amsterdam and Delta Institute for Theoretical Physics,\\ University of Amsterdam, Science Park 904, 1098 XH Amsterdam, The Netherlands}

\newcommand{\ICL}{Imperial College London, Exhibition Rd, South Kensington, London SW7 2BX, United Kingdom}

\begin{document}

%---- Title
\title{Perspectives on fundamental cosmology from Low Earth Orbit and the Moon}

\author{Gianfranco Bertone}
\email{g.bertone@uva.nl}
\affiliation{\GRAPPA}

\author{Oliver L. 
Buchmueller}
\email{o.buchmueller@imperial.ac.uk}
\affiliation{\ICL}

\author{Philippa S. Cole}
\email{p.s.cole@uva.nl}
\affiliation{\GRAPPA}
%\date{October 2021}

\begin{abstract}
    The next generation of space-based experiments will go hunting for answers to cosmology's key open questions which revolve around inflation, dark matter and dark energy. Low earth orbit and lunar missions within the European Space Agency's Human and Robotic Exploration programme can push our knowledge forward in all of these three fields. A radio interferometer on the Moon, a cold atom interferometer in low earth orbit and a gravitational wave interferometer on the Moon are highlighted as the most fruitful missions to plan and execute in the mid-term.
\end{abstract}


\maketitle

\section*{Introduction}
The standard cosmological model provides a simple framework to explain a variety of observations, ranging from sub-galactic scales to the size of the observable universe. Yet many open questions remain: the model relies on an unknown mechanism for the production of perturbations in the early universe, on an unknown matter component, generically referred to as dark matter, and on an unknown mechanism that leads to an accelerated expansion of the universe, generically referred to as dark energy.

The next generation of space-based experiments are our best chance of unveiling these mysteries. A united front of low earth orbit and lunar missions, as outlined in the European Space Agency's (ESA) Human and Robotic Exploration (HRE) \cite{agenda2025}, will break unprecedented ground on all of these fronts. Alongside the Laser Interferometer Space Antenna \cite{LISA_17}, a radio interferometer on the Moon, a cold atom interferometer in low earth orbit and a gravitational wave interferometer on the Moon would provide a full-coverage approach to unravelling the key open questions in cosmology today. 

%With many complementarities between observatories, the cohesive programme has the potential to make dramatic and perspective-changing discoveries.

In section \ref{sec:gaps} the key knowledge gaps in cosmology are highlighted, in section \ref{sec:prior} specific suggestions for experiments that should be the priorities for ESA's space programme and which questions they will answer are laid out, before concluding and discussing the future outlook in section \ref{sec:conc}.

\section*{Key knowledge gaps}\label{sec:gaps}

\subsection*{Inflation}

The theory of inflation is arguably the most promising model of the physics of the early universe \cite{PhysRevD.23.347}. The paradigm postulates that quantum fluctuations went on to seed the cosmological perturbations that we see imprinted on the Cosmic Microwave Background (CMB) and were the beginnings of all of the structure in the universe today. And yet, much remains to be understood about the properties of the quantum field that supposedly led to the initial period of exponential expansion of the universe. Whilst the paradigm is fully consistent with cosmological data \cite{Kogut_2003,Hinshaw_2003,Planckinflation}, we still currently lack direct smoking-gun evidence supporting it, as well as a specific model for how one or more scalar fields drove the expansion. 

On large scales, $k\sim10^{-3} – \,0.1\,{\rm Mpc^{-1}}$, observations of the CMB temperature anisotropies by Planck \cite{Planckinflation} have confirmed to incredible precision that density perturbations were small (fluctuations of order $10^{-5}$) and almost scale-invariant. The simplest single-field, slow-roll models of inflation are able to describe this spectrum of the density perturbations. However, deviations from scale-invariance on small scales could indicate a more complicated model that exhibits a feature in the inflationary potential. Such models could have interesting observational signatures, such as ultra-compact mini-haloes \cite{Aslanyan_2016,Bringmann_2012} or primordial black holes \cite{Garc_a_Bellido_2017}. Furthermore, primordial non-Gaussianity has been constrained to be small, $f_{\rm NL,local} = -0.9 \,\pm\, 5.1$, on large scales \cite{2020A&A...641A...9P}. This constraint has limited the viability of many models of inflation that predicted larger values of primordial non-Gaussianity, for example DBI inflation and EFT inflation \cite{Chen_2010,Renaux_Petel_2015}. However, reaching the $f_{\rm NL, local}<1$ threshold will provide strong evidence that observations are not consistent with multi-field models of inflation \cite{Byrnes_2010}. The final piece of the puzzle can be provided by the tensor-to-scalar ratio, which is currently constrained to be less than 0.1 \cite{Planckinflation}, a measurement of which would indicate the energy scale at which inflation happened.

\subsection*{Dark matter}

Similarly, the existence of dark matter is supported by a wide array of independent observations, but we still know very little about the fundamental nature of this elusive component of the universe. In the past four decades, a strong effort went into the search for a particular class of candidates: weakly interacting massive particles \cite{Schumann_2019}. However, no experiment has yet found evidence for these particles, and attention has turned to different classes of dark matter candidates in regions of parameter space where they would have evaded \textcolor{black}{strong constraints from direct detection} before now, for example axion-like-particles (ALPs) \cite{Choi_2021,ringwald,Irastorza:2018dyq} or primordial black holes (PBHs) \cite{Bertone_2018}. 

\textcolor{black}{Axion-like-particles are in particular a popular dark matter candidate \cite{Choi_2021,ringwald,Irastorza:2018dyq}. The QCD (quantum chromodynamics) axion was first postulated in the 70s to solve the strong CP problem \cite{axionellis}. However, ALPs more generally, often motivated by string theories in which ultra-light particles are ubiquitous, display the qualities required to explain all or part of the dark matter.} Whilst searches for the standard axion with a mass of order of a few hundred keV have yielded no detections, “invisible” axions with very small masses are still viable candidates. Search strategies vary depending on the mass of the axion, which can’t be theoretically predicted, but the most common approach is to probe their interactions with electromagnetic fields and constrain the axion-photon coupling \cite{axionhowtosee}. Astrophysical observations are able to look for signatures of axion to photon conversion in the presence of electromagnetic fields, for example, by looking for such processes in the vicinity of the magnetosphere of neutron stars \cite{Witte_2021,2018JHEP...06..036H,foster-witte}, or their production in the solar core, triggered by X-rays scattering off electrons and protons in the presence of the Sun’s strong \textcolor{black}{magnetic} fields \cite{1998PhLB..434..147M}.

For masses less than 1eV, axions are a sub-set of the broader class of ultra-light dark matter models, with masses down to (theoretically) $10^{-24}\,{\rm eV}$, although Lyman-alpha forest constraints have ruled out axion masses less than $2\times10^{-20}\,\mathrm{eV}$ \cite{Rogers_2021}, see \cite{uldm_review} for a review. \textcolor{black}{Ultra-light dark matter models} postulate a new ultra-light boson, which displays wave-like properties on galactic scales, but behaves like cold dark matter on larger scales where the \textcolor{black}{cold dark matter (CDM)} paradigm has strong support from observations. The behaviour on galactic scales, due to the Bose-Einstein condensate which forms, can have interesting signatures that could explain small-scale problems with CDM \cite{Del_Popolo_2017,Bullock_2017} and would have distinctive features for distinguishing between models such as fuzzy dark matter, self-interacting fuzzy dark matter and superfluid dark matter \cite{2017MNRAS.465..941D,Khoury_2022}.

Another promising candidate that received a lot of renewed attention after the LIGO/Virgo observations of order 10 solar mass binary black hole mergers is primordial black holes \cite{Bird:2016dcv,Sasaki_2016,Sasaki_2018}. They are the only proposed explanation of dark matter that requires no new physics beyond the standard model, which makes them an attractive candidate. However, strong constraints have now been placed across the parameter space via microlensing, gravitational waves and CMB observations which have essentially ruled them out as making up the entirety of the dark matter budget in all but one window around an asteroid mass. See e.g. \cite{Green:2020jor} for a review of current constraints. There is also the possibility of two-component dark matter models that include primordial black holes and another particle, with interesting signatures of interaction between the two \cite{Adamek_2019,Bertone_2019}.

\subsection*{Dark energy}
Dark energy is a generic term for the mechanism responsible for the observed accelerated expansion of the universe. One of the key questions that may bring us closer to the identification of dark energy is whether its energy density has remained constant throughout the history of the universe, as would be the case if it arises from the so-called vacuum energy, or whether it evolves with time, as appropriate for an evolving quantum field. See e.g. \cite{Mortonson_14} for a review.

A diversified experimental approach involving astronomical surveys and gravitational waves searches are arguably our best hope to make progress in the search for smoking-gun evidence of inflation, identification of dark matter, and understanding of dark energy. We will highlight possibilities for future space-based experiments that can break new ground on these frontiers.

\section*{Priorities for the space programme}\label{sec:prior}

Space experiments may soon provide important clues on the nature of inflation, dark matter and dark energy. As an overview of the current context, we list in Table 1 some experiments that might in particular enable gravitational wave searches for signatures of dark matter and primordial gravitational waves with space-borne interferometers
as well as indirect detection of dark matter and probing primordial fluctuations with Moon-based radio telescopes. We choose to focus on three key probes as most relevant in the framework of ESA’s Human and Robotic Exploration Directorate \cite{agenda2025} to address the knowledge gaps discussed above:

\begin{enumerate}[label=\Alph*)]
\item a radio interferometer on the Moon (RIM)
\item a space gravitational wave detector using cold atoms (AEDGE)
\item a gravitational waves interferometer on the Moon (GWIM)
\end{enumerate}

\begin{table*}
\resizebox{\textwidth}{!}{%
\begin{tabular}{|l|l|l|l|}
\hline
\textbf{Open fundamental scientific question} & \textbf{\begin{tabular}[c]{@{}l@{}}Focus of ESA HRE platform : \\ LEO, Moon, Mars, BLEO\end{tabular}} & \textbf{\begin{tabular}[c]{@{}l@{}}Context of related recent and \\ future space experiments\end{tabular}} & \textbf{Short, middle or long term} \\ \hline
Origin of primordial fluctuations             & Moon                                                                                                  & LISA, AEDGE, RIM                                                                                           & Middle term                         \\ \hline
Nature of dark matter                         & LEO                                                                                                   & LISA, AEDGE, AEDGE pathfinder, RIM, GWIM                                                                   & Middle term                         \\ \hline
Phase transitions                             & LEO                                                                                                   & LISA, AEDGE                                                                                                & Middle term                         \\ \hline
Existence of primordial black holes           & Moon                                                                                                  & LISA, AEDGE, GWIM                                                                                          & Middle term                         \\ \hline
\end{tabular}%
}
\caption{Recommendations for addressing key questions in fundamental cosmology with the ESA HRE programme in the short, middle and long term.}
\label{tab:HRE}
\end{table*}

\subsection*{Radio interferometer on the Moon}

The most relevant experiment for ESA's Directorate of Human and Robotic Exploration is the radio interferometer on the Moon \cite{Aminaei,2019arXiv190306212F, darkcosmo,Duke_85}. The rationale for this experiment is that placing a radio telescope on the far side of the Moon would give it access to wavelengths shorter than 30 MHz. Radiation at these frequencies is distorted or completely absorbed by the Earth’s ionosphere. An interferometer on the Moon will bypass this limitation, as well as shield the instruments from the background generated by terrestrial radio sources. Furthermore, the size of the array, which determines the resolution of the detector, is less restricted than an Earth-based detector like the Hydrogen Epoch of Reionization Array (HERA) \cite{DeBoer_2017}, the Low Frequency Array (LoFAR) \cite{refId0} or the Square Kilometre Array (SKA) \cite{SKAtel}. 

A radio telescope on the Moon would allow us to peer into the so-called dark ages of the universe \cite{Koopmans_19}, i.e. the epoch between the emission of the CMB and the reionization of the universe, triggered by the formation of the first stars. By studying the redshifted 21-cm line absorption feature, we can obtain unprecedented information on the history of reionization, and search for the signatures of dark matter annihilation or decay by looking for specific forms of the absorption feature \cite{Valdes_2007,Short_2020}.

Furthermore, a radio interferometer on the Moon could in principle have baselines as long as 300km, which would enable unprecedented access to information about very small scales \cite{Bernal_2018,Cole:2019zhu,Mena:2019nhm}. Measuring the 21cm power spectrum would provide a tracer for the underlying dark matter power spectrum. This could unlock information about dark matter sub-structures, the existence of primordial black holes, and the validity of slow-roll inflationary models. Tomographic analysis will also enable information to be gathered at a range of redshifts, providing new insight into the evolution of our Universe between the time of the CMB and today. 

Finally, a lunar radio telescope would allow us also to efficiently probe the so-called primordial non-gaussianity via the 21cm bispectrum \cite{Pillepich_2007,Mu_oz_2015,PNG_whitepaper}, which would provide a powerful test of the theory of inflation. 21cm observations will complement upcoming large-scale structure surveys which will help us to understand how structures are evolving, allowing for even more information to be extracted from the as yet un-probed region both in terms of redshift and access to small scales. 21cm observations from the Moon therefore add to the line of inquiry on all three fronts: inflation, dark matter and dark energy.

Arguably the largest challenge to overcome will be how to deal with extremely large foregrounds. They are expected to be 6 or 7 orders of magnitude larger than the signal being sought, and therefore systematics will need to be incredibly well understood. Extremely careful subtraction of galactic foregrounds \cite{Bernardi_2009, Bernardi_2010} will need to be performed so that any signal found in the data can be confidently interpreted \cite{Pober_2016,Makinen_2021,Liu_2009,Makinen_2021}.

\subsection*{Cold atom interferometer}

The direct detection of gravitational waves (GW) with LIGO/Virgo, that led to the Nobel prize for Physics in 2017, has opened new opportunities for cosmology. The space interferometer LISA has been selected to be ESA's third large-class mission, and it is scheduled to be launched in 2034. Experience with the electromagnetic spectrum shows the importance of measurements over a range of frequencies, and we note that there is a gap between the frequencies covered by LIGO/Virgo, as well as proposed detectors Einstein Telescope and Cosmic Explorer on Earth, and LISA in space, in the deci-Hz band. 

A promising candidate to explore this frequency band is the AEDGE mission concept~\cite{El_Neaj_2020}, which is based on novel quantum technology utilising Cold Atom (CA) techniques. Gravitational waves alter the distance between the cold atom clouds as they pass through the interferometer.

%to reach the required sensitivity to fully explore gravitational waves and dark matter searches in the deci-Hz band.
%AEDGE is a cold atom interferometer which has been proposed to search for ultra-light dark matter and gravitational waves simultaneously \cite{El_Neaj_2020}. 
%{\bf GB: I am not sure I understand the structure of this sentence:}  {\bf GB: Also, what to we mean with "required sensitivity to fully explore"?}
%Ultra-light bosonic fields would interact with the cold atoms and induce a shift in the relative phase between cold atom clouds by altering the frequency of the atomic transition being studied, whereas gravitational waves passing through the interferometer would induce a phase shift by altering the distance between the cold atom clouds. 

%Gravitational waves alter the distance between the cold atom clouds in the interferometer as they pass through in the deci-Hz band. These frequencies lie in a gap between LISA, ESA’s third large-class mission scheduled to launch in 2034, and ground-based detectors such as LIGO/Virgo . 

AEDGE will be able to probe the gravitational waves due to coalescing intermediate mass black holes with masses $100-10^5\,M_\odot$. This could shed light on the existence of intermediate black holes, and their potential role as seeds for the growth of supermassive black holes \cite{Latif_2016}. Additionally, black holes in the pair-instability mass gap will enter the deci-Hz range. If this gap is populated then it will motivate a deeper understanding of supernova collapse, or the need to invoke alternative mechanisms for producing such large black holes such as hierarchical mergers \cite{Di_Carlo_2020,Mehta_2022,Gerosa_2021}. Improved constraints on order $100 M_\odot$ primordial black holes should also be possible, which would complement constraints from the CMB on this mass range \cite{Poulin_2017}.

AEDGE can also probe a wide array of dark matter candidates. Scalar field dark matter, for instance, causes quantities such as the electron mass and the fine structure constant to oscillate with frequency and amplitude determined by the dark matter mass and local density. This leads to variation in the atomic transition frequencies, which imprint on the relative phase difference between cold atom clouds that atom interferometers measure. 

Proposed AEDGE sensitivities will enable the coupling between scalar dark matter and electrons, photons or via the Higgs-portal to be probed with up to 10 orders of magnitude improvement, with respect to current constraints from MICROSCOPE \cite{Berge_2017}, on mass ranges between $10^{-18}$ and $10^{-12}$eV. 
Other couplings can also be probed, for example the axion-nucleon coupling for axion-like DM lighter than $10^{-14}$eV, or the coupling between a dark vector boson and the difference between baryon and lepton number \cite{2016PhRvD..93g5029G,Poddar_2019}.

Furthermore, AEDGE offers a new channel for probing strong gravity regimes, where any deviations from general relativity are most likely to be noticeable. For example, precise measurements of post-Newtonian parameters would offer powerful probes of the predictions of general relativity and searches for deviations due to, for example, a graviton mass \cite{Will_1998}. Comparing measurements in different frequency ranges will also make possible sensitive probes of Lorentz invariance \cite{KOSTELECKY2016510}.

An opportunity to learn about the nature of dark matter from gravitational wave signals can also be realised by observations of coalescing black holes in the deci-Hz band. Intermediate mass black holes ($10^3– 10^5 M_\odot$) and primordial black holes (of any mass) may be surrounded by dense dark matter spikes \cite{Eda_2013,Eda_2015}. If these black holes form binaries with a much lighter companion, i.e. intermediate and extreme mass ratios ($q=m_2/m_1<10^{-2.5}$), then the dark matter may imprint a dephasing on the gravitational waveform of the inspiralling binary. The amount of dephasing could teach us something about the nature of the dark matter surrounding the black holes, for example whether it’s cold and collisionless, or whether it’s an ultra-light scalar field. LISA will have sensitivity to such effects for larger mass systems \cite{Kavanagh_2020,Coogan:2021uqv,Ng_2020}, future ground-based gravitational wave observatories will have sensitivity to low-mass systems, and an experiment such as AEDGE could bridge the gap to cover the entire frequency range and promote multi-band searches for the same signal. The gravitational wave signal could also be accompanied by an electromagnetic counterpart that might provide information about an environment involving dark matter if the particle is able to convert to radiation or interact with it in some way like it does in the case of, for example, axions.

Possible non-black hole binary cosmological targets for the deci-Hz band include GWs from first-order phase transitions in the early Universe, e.g., during electroweak symmetry breaking in modifications of the Standard Model with additional interactions, or during the breaking of higher gauge symmetries \cite{Zhou_2020}. The deci-Hz frequency range could also probe different parameter ranges for such transitions, and combining measurements with those by other experiments like LISA or LIGO/Virgo could help unravel different contributions, e.g., from bubble collisions, sound waves and turbulence \cite{Jinno_2017,Galtier_2021}. Another cosmological target for the deci-Hz band is the possible GW spectrum produced by cosmic strings \cite{Zhou_06}. In standard cosmology this spectrum would be almost scale-invariant, but there could be modifications due to a non-standard evolution of the early Universe. A detection or constraint on such observables could additionally provide clues as to the nature of dark matter, especially in the case of ultra-light dark matter particles which are expected to be produced alongside GWs from phase transitions in the early Universe and topological defects such as cosmic strings.

As outlined in the Cold Atom in Space Community Roadmap ~\cite{CW}, AEDGE, its pathfinder experiments, and other cold-atom experiments in space would also be able to make sensitive measurements relevant to several other aspects of fundamental physics, including the gravitational redshift, the equivalence principle, possible long-range fifth forces \cite{Battelier_2021,Elder_2016}, variations in fundamental constants and popular models of dark energy \cite{Sabulsky_2019}.

With the AION experiment in the UK~\cite{AION}, the MAGIS experiment in the US~\cite{MAGIS}, the MIGA experiment in France~\cite{MIGA}, the ZAIGA experiment in China~\cite{ZAIGA}, as well as the proposed European ELGAR project~\cite{ELGAR2}, there is already a large programme of terrestrial cold atom experiments in place. These experiments serve as terrestrial pathfinders for a large-scale mission like AEDGE, and it would be important to complement those with a dedicated technology development programme to pave the way for space-based cold atom pathfinder experiments.  First, dedicated pathfinders could be hosted at the International Space Station, building the foundation of a medium-class mission. This could then lead in the long-term to a large-class mission such as AEDGE to explore the ultimate physics potential of the deci-Hz band.      

\subsection*{A gravitational waves interferometer on the Moon}
In addition to a cold atom interferometer such as AEDGE, a lunar-based gravitational wave (GW) interferometer is ideal for probing GW frequencies in the range between deci-Hz to 5 Hz, which is where both Earth- and space-based detectors have less sensitivity \cite{1992LPI....23..751L,Jani_2021}. Preliminary estimates suggest that such an instrument would allow us to trace the expansion rate of the universe up to redshift $z\sim3$, test General Relativity and the standard cosmological model up to redshift $z\sim100$, and probe the existence of primordial black holes, as well as dark matter in neutron star cores \cite{PhysRevLett.126.141105,Dasgupta_2020}.

\section*{Future outlook and summary}\label{sec:conc}
We have outlined recommendations for how the future experiments in ESA's Human and Robotic Exploration Directorate fit into a strategy for answering the key fundamental questions in cosmology that we have highlighted in table \ref{tab:HRE}. These experiments will operate from either low earth orbit or the Moon, and are ambitious plans which should take place in the middle term, to give time for pathfinders and terrestrial experiments to pave the way for the development of the necessary technologies to make these moonshot missions possible. With many complementarities between experiments, European Space Agency’s (ESA) Human and Robotic Exploration programme has the potential to make dramatic and perspective-changing discoveries in cosmology and astroparticle physics.
%\pc{Final concluding remarks...}

\section*{Acknowledgements}

The authors thank the European Space Agency for the opportunity to contribute this review article. P.C. acknowledges funding from the Institute of Physics, University of Amsterdam. 

\subsection*{Competing interests}
The authors declare no competing financial or non-financial interests.

\subsection*{Author contributions}
Gianfranco Bertone and Oliver Buchmueller wrote the ESA science community white paper related to this topic, which Philippa Cole adapted to form this perspective piece.
\bibliography{main.bib}

\end{document}